# Pairing symmetries of several families of iron-based superconductors and some similarities with cuprates and heavy-fermions

Tanmoy Das[1]

[1]Theoretical Division, Los Alamos National Laboratory, Los Alamos, New Mexico 87544, USA

**Abstract.** We show that, by using unit-cell transformation between 1 Fe per unit cell to 2 Fe per unit cell, one can qualitatively understand the pairing symmetry of several families of iron-based superconductors. In iron-pnictides and iron-chalcogenides, the nodeless $s^\pm$-pairing and the resulting magnetic resonance mode transform nicely between two unit cells notations, while retaining all physical properties unchanged. However, when the electron-pocket disappears from the Fermi surface with complete doping in $KFe_2As_2$, we find that the unit-cell invariant requirement prohibits the occurrence of $s^\pm$-pairing symmetry (caused by inter-hole-pocket nesting). However, the intra-pocket nesting is compatible here, which leads to a nodal $d$-wave pairing. The corresponding Fermi surface topology and the pairing symmetry are similar to Ce-based heavy-fermion superconductors. Furthermore, when the Fermi surface hosts only electron-pockets in $K_yFe_{2-x}Se_2$, the inter-electron-pocket nesting induces a nodeless and isotropic $d$-wave pairing. This situation is analogous to the electron-doped cuprates, where the strong antiferromagnetic order creates similar disconnected electron-pocket Fermi surface, and hence nodeless $d$-wave pairing appears. The unit-cell transformation in $K_yFe_{2-x}Se_2$ exhibits that the $d$-wave pairing breaks the translational symmetry of the 2 Fe unit cell, and thus cannot be realized unless a vacancy ordering forms to compensate for it. These results are consistent with the coexistence picture of a competing order and nodeless $d$-wave superconductivity in both cuprates and $K_yFe_{1.6}Se_2$.

## 1 Introduction

In superconductors, the key process that allows current to travel without resistance is the formation of electron pair that moves as a single quantum state. The mechanism of pairing in all classes of unconventional superconductors including cuprates, pnictide, chalcogenides, heavy-fermion systems is still elusive. However, there are number of observations and theories that seem to be consistent across these materials, allowing us to make a comparative study and thereby, extract fundamental properties that are relevant to understand and formulate the mechanism of superconductivity. (1) *Dimensionality*: The unconventional superconductivity in cuprates, [1,2] iron-based compounds[3-6] and heavy-fermion [7,8] families occurs in their layered crystal structures. Moreover, it has been shown, at least in some crystal structures, that the value of $T_c$ increases as the number of superconducting (SC) layer is increased, and/ or the $c$-axis lattice constant is increased in cuprates,[1,2] pnictides[9] and heavy-fermions[10]. In particular, the value of $T_c$ increases almost linearly with the $c/a$ ratio [7,10] (where $a$ and $c$ are the in-plane and out-of-plane lattice constant of their common tetragonal structure), and also with the spin-fluctuation temperature $T_0$ (Ref. [11]). Notably, $c/a$ and $T_0$ have been argued to be interrelated.

(2) *Quantum critically*: All these materials consistently exhibit a dome-like behaviour of $T_c$ as a function of doping, or pressure or magnetic field. The maximum value of $T_c$ occurs in the vicinity of a quantum critical point of the normal state antiferromagnetic (AFM) or spin-density wave (SDW) order in cuprates,[12-17] Fe-based compounds, [18-22] and heavy-fermions [23-27]. There are numerous theoretical proposals suggesting that the critical fluctuations that drive the second order phase transition, also mediate or at least enhance pairing in these superconductors. [12-27] (3) *Spin-resonance mode*: The appearance of a strong magnetic mode at a characteristic spin-fluctuation wavevector $Q$ has been observed consistently in cuprates,[28-32] pnictides,[33-38] and heavy-fermion[39-42]. The weak or intermediate coupling theories have shown that a sign-reversal of the underlying SC order parameter between the Fermi momenta connected by the wave vector $Q$, can explain the occurrence of this mode.[22,27,43-53] Such a link between the Fermi surface (FS) topology and the sign-reversal SC gap seems to be consistent across all these materials,[32,53] indicating that the development of the magnetic mode, in turn, mediates Cooper pairing in repulsive interaction environment.

In this paper, we mainly focus on the third scenario; however also comment on the relevant first and second pictures for these families of superconductors. We point out some of the subtleties involved in deriving the link between the FS topology and sign-changing pairing for multiband systems. As a starting point, we take a FS which consists of two disconnected pockets situated at $\Gamma(0,0,0)$ and $M(\pi,\pi,0)$ momenta. This is the typical FS topology for iron-pnictide and iron-chalcogenides (single-layer) superconductors. Mazin *et al.* [47] have shown that the interband nesting between these pockets gives rise to a spin-resonance mode if the SC gap on each band is isotropic but changes sign between them. This scenario predicts a so-called $s^{\pm}$-pairing symmetry. In real-space, the $s^{\pm}$-pairing leads to opposite SC phases between the two iron atoms sitting at the corner and the center of the Brilouin zone. Theories constructed based on 1 Fe atom per unit cell (hence say 1FUC) or 2 Fe atoms per unit cell (hence say 2FUC) obtain the same physical phenomena, given that one performs the same unitary transformation on the Fermi surface as well as on the pairing symmetry. [22,54,55]

The situation becomes more complicated and exotic when the FS pocket either at $\Gamma$ or at M point completely disappears. The aforementioned theory suggests that due to the lack of interband nesting to produce spin fluctuations, superconductivity should vanish. However, the material realization of both these situations with considerably large value of $T_c$ in various iron-based superconductors have revamped our general consensus. We here show that the spin-fluctuation theory still holds, if the pairing symmetry changes accordingly to maintain the sign-reversal property. We propose here that the unitary transformation between 1FUC↔2FUC is a simple, yet very efficient tool to uncover the pairing symmetry that should be compatible with the lattice geometry, and also can given rise to a spin resonance. We will not perform any numerical calculation to test which pairing state is energetically favourable for a particular situation, but only provide the symmetry arguments for the possible pairing state.

Our results include: (1) In iron-pnictide, the full hole-doping completely eliminates the electron pockets from the M points in $KFe_2As_2$. The intra-hole-pocket nesting can, in principle, produce some form of $s^{\pm}$-pairing that will continue to maintain both sign-changing and isotropic gap features. However, we show that the requirement for the invariance of the physical properties such as magnetic resonance under the 1FUC↔2FUC unitary transformation is violated for this pairing state. On the other hand, an anisotropic and nodal $d$-wave pairing is consistent with all these properties and can also explains numerous experimental data in $KFe_2As_2$.[56-60] Considering the similarity of the FS topology between $KFe_2As_2$ and heavy-fermion superconductor $CeCoIn_5$, we conjecture that the latter also hosts nodal $d$-wave pairing which leads to spin-resonance, consistent with experiment.[40] (2) The counter example of the above situation is also realized in double layered iron-chalcogenide superconductors.[4-6] In this case, the electron-pocket at M point is eliminated by increasing number of iron-vacancies in the non-stoichiometric $K_yFe_{2-x}Se_2$ crystal.[61-63] Both theories [22,50,52] and experiments [64-66] have demonstrated that isotropic and nodeless $d$-wave pairing is energetically favourable in this case, originating from intra-electron-pockets nesting. Our theoretical prediction of the spin-resonance [22] as a result of such nesting, has been observed recently by inelastic neutron scattering measurement.[67] The other possibilities of nodeless pairing such as $s^{\pm}$ or $s^{++}$ or $s^{--}$, [68,69] or even nodal pairings [70] may be incompatibility when performing the unit cell transformation including Fe vacancies.[54] Although, the nodal planes are present in the underlying $d$-wave pairing state, but they do not intersect the FSs, and thus the low-energy quasiparticles exhibit fully gapped density of states at the Fermi level. The induction of nodeless and isotropic pairing on the FS, even when the underlying pairing symmetry is nodal $d$-wave is remarkably analogous to the electron-doped cuprates in their underdoped regime. In the latter case, the strong AFM ordering shrinks the large FS into small and disconnected electron pockets at M points, and thus nodeless $d$-wave pairing appears.[71,72]

## 2 Unit-cell transformation in 122-crystal

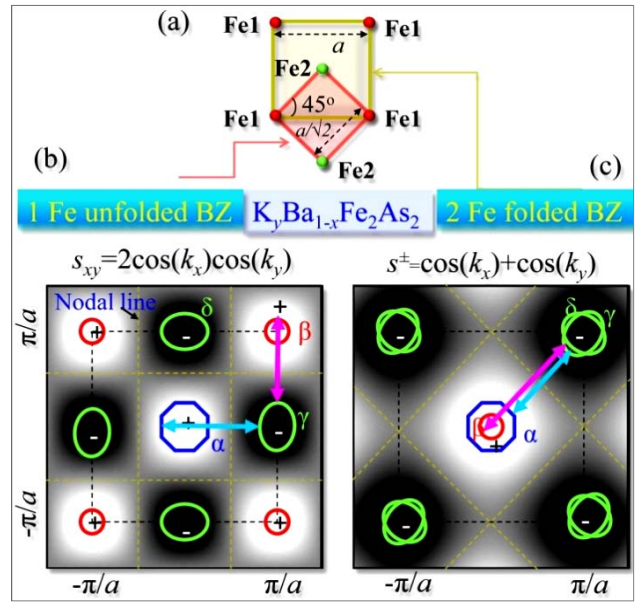

**Fig. 1.** (a) Unit-cell transformation between 1FUC↔2FUC (schematic). (b) FS topology of $K_xBa_{1-x}Fe_2As_2$ superconductors (schematic) in the 1FUC. The black (negative) to white (positive) background represents the $s^{\pm}$-pairing symmetry. (c) Same as (b), but in the 2FUC.

Ba-based 122 pnictide superconductor, $K_xBa_{1-x}Fe_2As_2$, belongs to the *I4/mmm* crystal symmetry in which two inequivalent Fe atoms reside at the corner and center of the Brillouin zone. For most of the practical purposes, these two Fe atoms are assumed to be equivalent, which allows us to construct a conventional two-dimensional unit cell containing only 1 Fe atom, see figure 1(a). The unitary transformation between 1FUC↔2FUC consists of $45^o$ rotation of the crystal with lattice constant $a{\rightarrow}a/\sqrt{2}$ which gives $k_{x/y}=(k_x{\pm}k_y)/2$. In doing so, the M point of the Brillouin zone lies at $(\pi,0)$ in 1FUC and at $(\pi,\pi)$ in 2FUC. And the pairing function is $s_{xy}=2\cos(k_xa)\cos(k_ya)$ and $s^{\pm}=\cos(k_x a)+\cos(k_y a)$ in 1FUC and 2FUC, respectively. In such transformation, the nodeless and isotropic natures of the pairing symmetry remain consistent in both cases. This situation can be understood by comparing figure 1(b) for 1FUC with figure 1(c) for 2FUC. The resulting magnetic resonance mode, which occurs due to the sign reversal of the SC gap at the hot-spot vectors [blue and magenta arrows in figures 1(b) and 1(c)] appear on both unit cell notations at the same energy, but aligned along the $q \rightarrow (\pi, 0)$ direction in the 1FUC and $q \rightarrow (\pi, \pi)$ direction in the 2FUC.[33-38,49] Furthermore, as $\alpha$ and $\beta$ hole pockets acquire different magnitude of SC gaps, [73,74] we have shown in a previous calculation [51] that the spin excitation spectrum splits into two resonance energy scales.

## 3 KFe$_2$As$_2$, no electron-pocket and nodal *d*-wave pairing

While the nodeless nature of the pairing symmetry has been consistent with several measurements [20,73,74] in K$_x$Ba$_{1-x}$Fe$_2$As$_2$, the experimental finding of nodal SC state in its extreme doping regime, i.e. in KFe$_2$As$_2$ contradicts this scenario. Bulk measurements including NMR, [56,57] thermal conductivity, [58,59] penetration depth [60] clearly demonstrate the presence of a linear-in-*T* dependence in their low-temperature behavior, which is taken as a clear demonstration of the nodal SC state. The nodal structure has also been indicated in LaFePO [75,76] and BaFe$_2$(As$_{1-x}$P$_x$) [77,78] systems.

At fully hole-doped KFe$_2$As$_2$, the electron pockets disappear from the FS.[73,79] Since two remaining hole-pockets are present at $\Gamma$ and $(\pi,\pi)$ points, one can argue that the $s^{\pm}$-pairing in the 1FUC convention [same to the pairing state of K$_x$Ba$_{1-x}$Fe$_2$As$_2$ in the 2FUC] can be stabilized in this case, according to the theory of sign-reversal SC pairing for the presence of a spin-resonance mode.[47-49] However, the requirement for the unitary transformation invariance between 1FUC$\leftrightarrow$2FUC contradicts this possibility. As we see in figure 2(b), both hole-pockets are now moved to $\Gamma$ point in the 2FUC, while the pairing symmetry switches to $s_{xy}$. As the sign-changing does not occur for either intraband or interband nesting in this unit cell convention (which is the actual unit cell for this material), a spin resonance is prohibited to appear in the SC state. On the basis of this reason, $s^{\pm}$- (or $s_{xy}$-) pairing is ruled out for this case.

Therefore, each pocket should host nodal lines that will pass through either the diagonal direction ($d_{x^2-y^2}$) or parallel to the zone boundary ($d_{xy}$). Figure 2(c) depicts the situation for $d_{x^2-y^2}$-wave pairing in the 1FUC which transforms properly to the 2FUC, and all physical properties such as nodal and anisotropic quasiparticles, resulting spin-resonance mode remain unchanged in both unit cell notations. The $d_{x^2-y^2}$–wave pairing in the 1FUC transforms to an extended $d_{xy}=2\sin(k_x a/2)\sin(k_y a/2)$ (not to be confused with the typical $d_{xy}$ pairing) in the 2FUC, and a spin-resonance mode will commence in the low-energy region near $q \sim (\pi,\pi)$ in the 1FUC and $q \sim (\pi,0)$ in the 2FUC. Previously, a renormalization group theory calculation[79] finds that a combination of $s^{\pm}$ [figure 2(a)] and $d_{x^2-y^2}$ [figure 2(c)] wave pairing which leads to an exotic nodal gap $[\cos(k_x a)+\cos(k_y a)][\cos(k_x a)-\cos(k_y a)]$, is energetically favorable in this case. It should be noted that in the case when $\beta$ pocket also disappears from the FS, $d_{x^2-y^2}$ pairing will still survive.

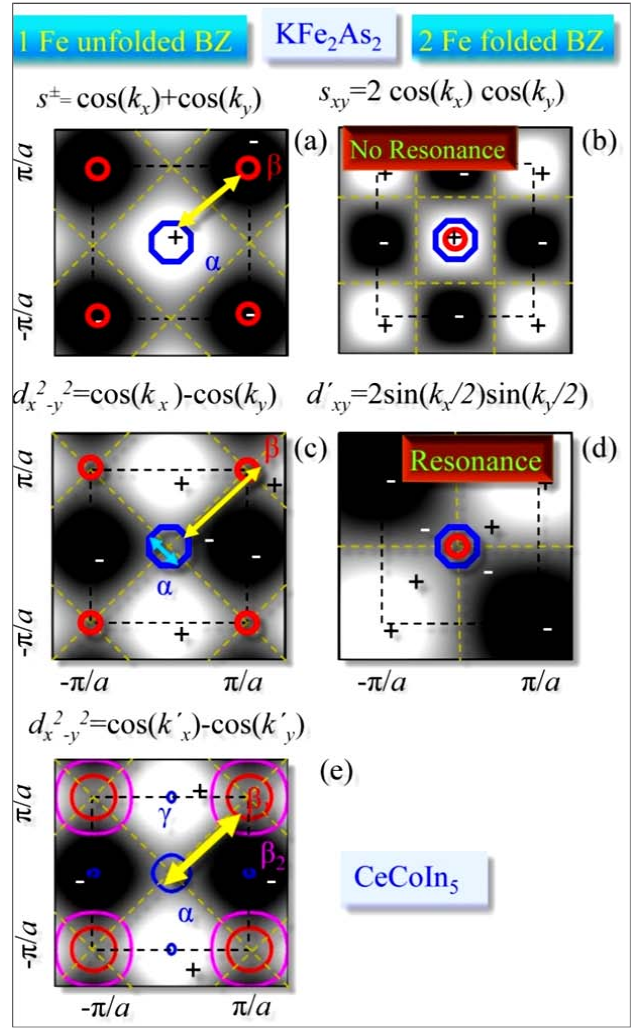

**Fig. 2. (a)** FSs, hosting only hole-pockets (schematic), and $s^{\pm}$-pairing symmetry for KFe$_2$As$_2$ system in 1FUC. **(b)** In 2FUC, both hole-pockets collapse at $\Gamma$ point and the pairing symmetry transformed to a $s_{xy}$-wave one. Due to the lack of sign-reversal of the SC gap on the FS, no resonance is expected here. **(c)-(d)**, Same as **(a)** and **(b)**, respectively, but for *d*-wave pairing state. Arrows depict all possible dominant nesting channels. Due to sign-reversal, a resonance mode can be expected here. **(e)** The computed FS and *d*-wave pairing for heavy-fermion CeCoIn$_5$ superconductor.

### 3.1 Similarity between KFe$_2$As$_2$ and CeCoIn$_5$ superconductors

The characteristic similarity of the FS topology between KFe$_2$As$_2$ and heavy fermion CeCoIn$_5$ system can be realized by comparing figures 2(c) and 2(e). The FS for CeCoIn$_5$ is derived from tight-binding parametrization to the first-principles band structure, [80-82] which matches well with angle-resolved photoemission spectroscopy (ARPES) study [83] as well as with Shubnukov de-Haas van Alphen measurement [84]. Here, multiple FS nestings along the diagonal direction dominate, while the nesting along $q \sim (\pi, 0)$ between $\alpha$ and $\beta_2$ FSs is naturally weaker. Therefore, it is viable to conclude that a $s^{\pm}$-wave pairing that appeared in $K_xBa_{1-x}Fe_2As_2$ for the strong nesting along $q \sim (\pi, 0)$ in figure 1, will be overturned by the $d_{x^2-y^2}$-pairing, as in $KFe_2As_2$. A resonance will commence for the $d_{x^2-y^2}$-pairing which will be aligned along the $q \to (\pi, \pi)$ direction.[46] Experimental confirmation of the presence of spin-resonance measurement [39-42] and nodal $d_{x^2-y^2}$-wave pairing [85-88] have been obtained extensively in this class of materials.

For PuCoGa$_5$ and other actinide superconductors, the hole pockets at M-points are larger (and also more number of concentric pockets are present here [27]). A $d_{x^2-y^2}$-wave pairing will thus be favorable, consistent with superfluid density calculations.[89] For cuprates, a large hole-like FS is only present centering M point, stipulating a intraband nesting at $Q \sim (\pi, \pi)$, and thus nodal $d_{x^2-y^2}$-pairing appears.[43-45]

# 4 $K_yFe_{2-x}Se_2$, no hole-pocket and nodeless $d$-wave pairing

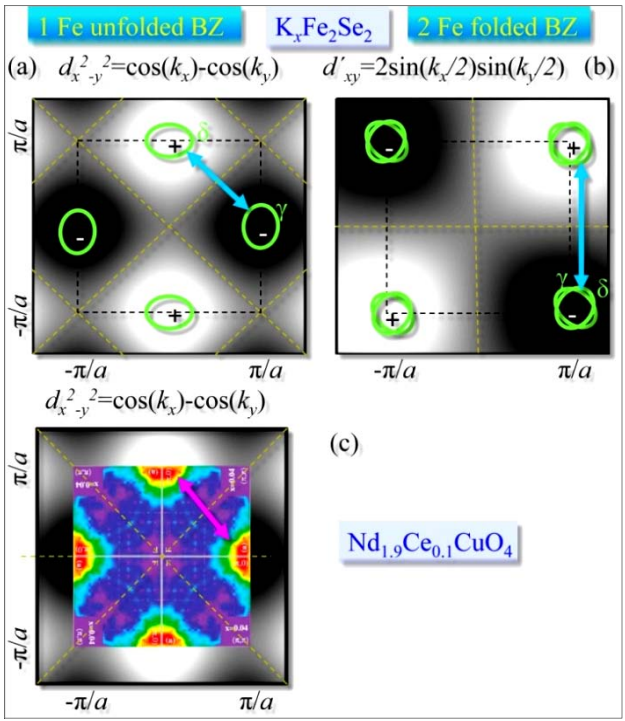

**Fig. 3. (a), (b)** Schematic FS and $d$-wave pairing for $K_xFe_{1.6}Se_2$ system in the 1FUC and 2FUC, respectively. **(c)** ARPES results of FS (symmetrized on the whole Brillouin zone) for an electron-doped superconductor. Strong AFM order in underdoping eliminates the hole-pocket from the nodal region, and thus a nodeless $d$-wave pairing emerges in this cuprate system as in $K_xFe_{1.6}Se_2$ system.

We now study the case in which there is no hole pocket present on the FS. In principle, we can achieve this situation in iron pnictide materials via heavy electron doping; however at such large doping superconductivity also disappears.[90] On the other hand, the recently discovered superconductivity in iron-selenide compound in the 122 crystal structure with $T_c \sim 37$ K makes the material realization of this situation possible.[4-6] In these materials, the number of Fe vacancies in the crystal acts as a tuning parameter for the electronic states. Above a certain value of Fe vacancies, the hole-pockets are removed from $\Gamma$ point, and the FS accommodates only electrons pockets at M points, as shown in figure 3(a).[61-63] The inter-electron-pocket nesting has been shown by numerous calculations to lead to a $d_{x^2-y^2}$ pairing.[22,50, 52] However, unlike the nodal $d$-wave in $KFe_2As_2$ discussed in Sec. 3, here it is nodeless and isotropic. The singlet, [91] nodeless, and isotropic nature of the pairing has been observed by numerous experiments.[64-66] Again, based on the sign-reversal properties of this $d$-wave gap, random-phase approximation (RPA)-based susceptibility calculation has predicted the presence of a spin-resonance mode,[22] which is observed later by inelastic neutron scattering measurement.[67]

The unitary transformation of the FS and the pairing symmetry to the 2FUC is more interesting. As seen in figure 3(b), both FS pieces are accumulated at $(\pi, \pi)$ point in the 2FUC and the extended $d_{xy}$ pairing in this unit cell leads to a sign change of the gap along $q \sim (2\pi, 0)$ direction. All the observables including nodeless, isotropic SC state, and magnetic resonance mode are preserved in both unit cell notations. However, the pairing symmetry in the 2FUC clearly breaks the translational symmetry. Therefore, such pairing can only be realized if the pairing state appears when the normal state already accommodates the broken translational symmetry. It has been shown by numerous experiments including NMR, [92] transmission electron microscopy (TEM), [93] neutron diffraction, [94] and optical spectra [95] that these materials become superconducting exactly when the randomly created Fe vacancies form an ordered state. We have shown in our previous calculation [54] that such vacancy ordering leads to a structural transition from the $I4/mmm$ structure to a lower-symmetry $I4/m$ structure (also called $\sqrt{5} \times \sqrt{5} \times 1$ unit cell), which is compatible with the $d$-wave pairing in any unit-cell notations. With a mean-field treatment, we have also demonstrated that the incommensurate nature of the vacancy order at $Q_v$ naturally promotes a charge density wave (CDW), orbital density wave (ODW) and spin-density wave (SDW) at $2Q_v$, $3Q_v$, $4Q_v$.[54] The modulation vectors $Q_v$, [96,97] $2Q_v$, [98] and $4Q_v$,[94] have been observed experimentally for the vacancy concentration of $K_yFe_{1.6}Se_2$, while $3Q_v$ has yet not been reported. A strong enhancement of magnetism at $4Q_v$ has been reported consistently to occur at the same time when vacancy order and superconductivity have turned on. [94,96] All these order parameters seem to coexist with $d$-wave pairing to lead to a 'modulated superconductivity' or 'staggered superconductivity'. In this case, all order parameters and superconductivity are tied to each other by symmetry considerations. Much like SO(5) symmetry for the AFM $d$-wave pairing [99] or SO(8) for `stripe' $d$-wave pairing [100] proposed for cuprates, in this iron-selenide compound the modulated pairing will obey a SO($n$) algebra, where the value of $n$ will depend on $Q_v$.

## 4.1 Similarity between $K_yFe_{1.6}Se_2$ and underdoped $Nd_{1.9}Ce_{0.1}CuO_4$

It is interesting to compare the FS topology of $K_yFe_{1.6}Se_2$ with that of electron-doped cuprate $Nd_{1.9}Ce_{0.1}CuO_4$. In the latter system, the strong AFM order breaks the fully metallic FS into electron pockets at $(\pi,0)$ and its equivalent momenta and hole-pocket at $(\pi/2,\pi/2)$. [101] When the strength of the AFM gap increases with underdoping, it eliminates the hole-pocket from the nodal region. And the FS is now left with only electron pockets, as observed by ARPES, [102] see figure 3(c). Therefore, the underlying $d_{x^2-y^2}$-pairing exhibits nodeless and isotropic feature, consistent with the exponential behavior of the penetration depth measurements.[71] This feature is generic to all electron-doped materials.[72] For this system, a magnetic resonance behavior is observed in inelastic neutron scattering measurement which is consistent with the same scenario of the sign-changing gap.[103] As doping increases and the AFM gap decreases, the hole-pocket appears at the nodal-point, and the nodal $d$-wave pairing is restored.[72] For hole-doping, hole-pocket is always present, and thus the nodal gap structure persists at all doping.[53,104]

## 5 Conclusions

We show that the unitary transformation between 1FUC↔2FUC can be a useful tool to understand the pairing symmetry of iron-based superconductors. For fully hole-doped doped 122 compound, $KFe_2As_2$, when electron-pocket disappears from the FS, we show that a $s^\pm$-paring coming from the inter-hole-pocket-nesting is ruled out by the unit cell transformation invariance requirement. However, a nodal $d_{x^2-y^2}$-wave pairing is most likely be the stable pairing state in this case. This scenario is analogous the Ce-based heavy-fermion superconductors. Furthermore, for $K_yFe_{1.6}Se_2$ systems, when the hole-pockets disappear, the inter-electron-pocket nesting gives nodeless and isotropic $d_{x^2-y^2}$-wave pairing. This situation is similar to the electron-doped cuprates, where the AFM correlation breaks the full FS into electron pockets with same topology as in $K_yFe_{1.6}Se_2$. In both cases, a transition to the nodeless $d_{x^2-y^2}$-wave pairing has been observed extensively. The 1FUC↔2FUC unitary transformation exhibits that the $d_{x^2-y^2}$-wave pairing breaks translational symmetry, and thereby requiring a normal state broken symmetry to compensate for it.[54] The result explains the coexistence of AFM and $d_{x^2-y^2}$-wave in cuprates and the same for vacancy order with $d_{x^2-y^2}$-wave in $K_yFe_{1.6}Se_2$ in an equal footing. This consistency also demonstrates the importance of quantum criticality for superconductivity in these systems.

## Acknowledgement

The author is grateful to A. V. Balatsky, M. J. Graf, J.-X. Zhu, and Nai-Chang Yeh for useful discussions. This work was supported by the U.S. DOE at Los Alamos National Laboratory under contract No. DE-AC52-06NA25396 and No. DE-AC52-06NA25396 through the LDRD program and the Office of Basic Energy Science (BES) and benefited from the NERSC computing allocations.

## References

1. T. Sato, *et al.* Phys. Rev. Lett. **89**, 067005 (2002).
2. C. Ambrosch-Draxl, *et al.* Phys. Rev. Lett. **92**, 187004, (2004).
3. J. Paglione, and R. L. Greene, Nature Physics **6**, 645-658 (2010).
4. J. Guo, *et al.* Phys. Rev. B **82**, 180520(R) (2010).
5. A. Krzton-Maziopa, *et al.* J. Phys. Condens. Matter **23**, 052203 (2011).
6. M. Fang, *et al.* Europhys. Lett. **94**, 27009 (2011).
7. P.G. Pagliuso, *et al.* Physica B **312**, 129 (2002).
8. J. L. Sarrao, *et al.* Nature **420**, 297-299 (2002).
9. N. Ni, J. M. Allred, B.. C. Chan, R. J. Cava, arXiv:1106.2111.
10. E. D. Bauer, *et al.* Phys. Rev. Lett. **93**, 147005 (2004).
11. N. J. Curro, *et al.* Nature **434**, 622-625 (2005).
12. Matthias Vojta, Rep. Prog. Phys. **66**, 2069 (2003).
13. Tanmoy Das, R. S. Markiewicz, and A. Bansil, Phys. Rev. B **74**, 020506(R) (2006).
14. S. Sachdev, Physica Status Solidi B **247**, 537 (2010).
15. E. G. Moon, S. Sachdev, Phys. Rev. B **82**, 104516 (2010).
16. Tanmoy Das, R. S. Markiewicz, and A. Bansil, Phys. Rev. B **81**, 184515 (2010).
17. Tanmoy Das, R. S. Markiewicz, and A. Bansil, Phys. Rev. B **81**, 174504 (2010).
18. H. Luetkens, *et al.* Nat. Mater. **8**, 305-309 (2009).
19. J. Dai, Q. Si, J.-X. Zhu, E. Abrahams, Proc. Nat. Acad. Sci. (PNAS), **106**, 4118-4121 (2009).
20. D. C. Johnston, Adv. in Phys. **59**, 803-1061 (2010), and the references therein.
21. E. Abrahams, Q. Si, J. Phys.: Condens. Mat. **23**, 223201 (2011).
22. Tanmoy Das, and A. V. Balatsky, Phys. Rev. B **84**, 014521 (2011).
23. J. Custers, *et al.* Nature **424**, 524-527 (2003).
24. J. R. Jeffries, *et al.* Phys. Rev. B **72**, 024551 (2005).
25. P. Gegenwart, Q. Si, and F. Steglich, Nature Physics **4**, 186-197 (2008).
26. T. Park, *et al.* Nature **440**, 65-68 (2006).
27. Tanmoy Das, Jian-Xin Zhu, and Matthias J. Graf, arXiv:1108.0272, Phys. Rev. Lett. (2011).
28. H. F. Fong, *et al.* Nature **398**, 588 (1999).
29. P. Bourges, *et al.* Science **288**, 1234 (2000).
30. S. D. Wilson, P. Dai, S. Li, S. Chi, H. J. Kang, and J. W. Lynn, Nature **442**, 59 (2006).
31. B. Vignolle, *et al.* Nature Physics **3**, 163 (2007).
32. G. Yu, Y. Li, E. M. Motoyama, M. Greven, Nature Physics **5**, 873 (2009).
33. A. D. Christianson, *et al.* Nature **456**, 930 (2008).
34. M. D. Lumsden, *et al.* Phys. Rev. Lett. **102**, 107005 (2009).

35. Y. Qiu, *et al.* Phys. Rev. Lett. **103**, 067008 (2009).
36. D. N. Argyriou, *et al.* Phys. Rev. B **81**, 220503(R) (2010).
37. S. Li, *et al.* Phys. Rev. Lett. **105**, 157002 (2010).
38. O. J. Lipscombe, *et al.* Phys. Rev. B **82**, 064515 (2010).
39. N. K. Sato, *et al.* Nature **410**, 340 (2001).
40. C. Stock, C. Broholm, J. Hudis, H. J. Kang, and C. Petrovic, Phys. Rev. Lett. **100**, 087001 (2008).
41. F Steglich, *et al.* J. Phys.: Condens. Matter **22** 164202 (2010).
42. B. Fak, S. Raymond, D. Braithwaite, and G. Lapertot, and J.-M. Mignot, Phys. Rev. B **78**, 184518 (2008).
43. Ar. Abanov, and A. V. Chubukov, Phys. Rev. Lett. **83**, 165 (1999).
44. M. Eschrig, and M. R. Norman, Phys. Rev. B **67**, 1445031(2003).
45. I. Eremin, *et al.* Phys. Rev. Lett. **94**, 147001 (2005).
46. The three dimensionality of the Fermi surface topology shift the resonance to $(\pi,\pi,\pi)$. [A. V. Chubukov and L. P. Gorkov, Phys. Rev. Lett. **101**, 147004 (2008).]
47. I. I. Mazin, D. J. Singh, M. D. Johannes, M. H. Du, Phys. Rev. Lett. **101**, 057003 (2008).
48. A. V. Chubukov, D. V. Efremov, and I. Eremin, Phys. Rev. B **78**, 134512 (2008).
49. T. A. Maier, S. Graser, D. J. Scalapino, and P. Hirschfeld, Phys. Rev. B **79**, 134520 (2009).
50. T. A. Maier, S. Graser, P. J. Hirschfeld, and D. J. Scalapino, Phys. Rev. B **83**, 100515(R) (2011).
51. Tanmoy Das, and A. V. Balatsky, Phys. Rev. Lett. **106**, 157004 (2011).
52. F. Wang, F. Yang, M. Gao, Z.-Yi Lu, T. Xiang, and D.-H. Lee, Europhys. Lett. **93**, 57003 (2011).
53. Tanmoy Das, R. S. Markiewicz, and A. Bansil, arXiv:1110.0756.
54. Tanmoy Das, and A. V. Balatsky, Phys. Rev. B **84**, 115117 (2011).
55. Tanmoy Das, and A. V. Balatsky, arXiv:1110.3834.
56. H. Fukazawa, *et al.* J. Phys. Soc. Jpn. **78**, 083712 (2009).
57. S. W. Zhang, *et al.* Phys. Rev. B **81**, 012503 (2010).
58. J. K. Dong, *et al.* Phys. Rev. Lett. **104**, 087005 (2010).
59. T. Terashima, *et al.* Phys. Rev. Lett. **104**, 259701 (2010).
60. K. Hashimoto, *et al.* Phys. Rev. B **82**, 014526 (2010).
61. Y. Zhang, *et al.* Nature Materials **10**, 273 (2011).
62. X.-P. Wang, *et al.* Europhys. Lett. **93**, 57001 (2011).
63. D. Mou, *et al.* Phys. Rev. Lett. **106**, 107001 (2011).
64. Z. Shermadini, *et al.* Phys. Rev. Lett. **106**, 117602 (2011).
65. H. Kotegawa, *et al.* Phys. Soc. Jpn. **80**, 043708 (2011).
66. D. A. Torchetti, *et al.*, Phys. Rev. B **83**, 104508 (2011).
67. J. T. Park, *et al.* arXiv:1107.1703.
68. I. I. Mazin, Phys. Rev. B **84**, 024529 (2011); Physics **4**, 26 (2011).

69. H. Kontani, and S. Onari, Phys. Rev. Lett. **104**, 157001 (2010).
70. R. Yu, P. Goswami, Q. Si, P. Nikolic, J.-X. Zhu, arXiv:1103.3259.
71. Tanmoy Das, R. S. Markiewicz, and A. Bansil, Phys. Rev. Lett. **98**, 197004 (2007).
72. Tanmoy Das, R. S. Markiewicz, and A. Bansil, J. Phys. Chem. Sol. **69**, 2963 (2008).
73. K. Nakayama, *et al.* Phys. Rev. B **83**, 020501 (2011).
74. M. L. Teague, G. K. Drayna, G. P. Lockhart, P. Cheng, B. Shen, H.-H. Wen, and N.-C. Yeh, Phys. Rev. Lett. **106**, 087004 (2011).
75. J. D. Fletcher, *et al.* Phys. Rev. Lett. **102**, 147001 (2009).
76. M. Yamashita, *et al.* Phys. Rev. B **80**, 220509(R) (2009).
77. K. Hashimoto, *et al.* Phys. Rev. B **81**, 220501(R) (2010).
78. Y. Nakai, *et al.* Phys. Rev. B **81**, 020503(R) (2010).
79. R. Thomale, *et al.* Phys. Rev. Lett. **107**, 117001 (2011).
80. Tanmoy Das, and Matthias J. Graf, unpublished.
81. T. Maehira, T. Hotta, K. Ueda, and A. Hasegawa, J. Phys. Soc. Jpn. **72**, 854-864 (2003).
82. H. Shishido, *et al.* J. Phys. Soc. Jpn. **71**, 162 (2002).
83. A. Koitzsch, *et al.* Phys. Rev. B **79**, 075104 (2009).
84. H. Shishido, *et al.* J. Phys.: Cond. Mat. **15**, L499 (2003).
85. N. Hiasa and R. Ikeda, Phys. Rev. Lett. **101**, 027001 (2008).
86. W. K. Park, J. L. Sarrao, J. D. Thomson, and L. H. Greene, Phys. Rev. Lett. **100**, 177001 (2008).
87. K. An, *et al.* Phys. Rev. Lett. **104**, 037002 (2010).
88. K. M. Suzuki, *et al.* J. Phys. Soc. Jpn. **79**, 013702 (2010).
89. Tanmoy Das, Jian-Xin Zhu, and Matthias J. Graf, Phys. Rev. B **84**, 134510 (2011).
90. Y. Sekiba, *et al.* New J. Phys. **11**, 025020 (2009).
91. W. Yu, *et al.* Phys. Rev. B **83**, 104508 (2011).
92. D. A. Torchett, *et al.* arXiv:1111.2552.
93. Z. Wang, *et al.* Phys. Rev. B **83**, 140505(R) (2011).
94. W. Bao, *et al.* Chinese Phys. Lett. **28**, 086104 (2011).
95. Z. G. Chen, *et al.* Phys. Rev. B **83**, 220507(R) (2011).
96. F. Ye, *et al.* Phys. Rev. Lett. **107**, 137003 (2011).
97. Z.Wang, *et al.* Phys. Rev. B **83**, 140505(R) (2011).
98. V. Yu. Pomjakushin, *et al.* Phys. Rev. B **83**, 144410 (2011).
99. Shou-Cheng Zhang, Science **275**, 1089 (1997).
100. R. S. Markiewicz, and M. T. Vaughn, Phys. Rev. B **57**, R14052 (1998).
101. C. Kusko, R. S. Markiewicz, M. Lindroos, and A. Bansil, Phys. Rev. B **66**, 140513 (2002).
102. N. P. Armitage, *et al.* Phys. Rev. Lett. **88**, 257001 (2002).
103. G. Yu, Y. Li, E. M. Motoyama, R. A. Hradil, R. A Mole, and M. Greven, arXiv:0803.3250.
104. Tanmoy Das, R. S. Markiewicz, and A. Bansil, Phys. Rev. B **77**, 134516 (2008).